\documentclass[aps, twocolumn]{revtex4}
\usepackage{amssymb}
\usepackage{amsmath}
\usepackage{epsfig}
\begin{document}

%__________________________________________________________________________
%
%%\documentstyle[preprint,aps,amssymb]{revtex}  % PREPRINT format REVTEX 4.0
%\documentclass[draft,twocolumn,aps]{revtex4} % TWOCOLUMN format REVTEX 4.0
%%\documentstyle[aps]{revtex}           % GALLEY format REVTEX 4.0
%option: preprint

                           % this command makes pacs numbers print

%-----------------------------------------------------------------------------

%\title{\vskip -0.5in\hfill\hfil{\rm\normalsize Printed on \today}\vskip 0.4in

\title{Magnetoresistance due to edge spin accumulation}                           

\author{M. I. Dyakonov}   

\affiliation{Laboratoire de Physique Th\'eorique et Astroparticules,
Universit\'e Montpellier II, CNRS, France}

%\date{Received \quad}

%-----------------------------------------------------------------------------

\begin{abstract}
 Because of spin-orbit interaction, an electrical current 
is accompanied by a spin current resulting in spin accumulation near the sample edges. 
Due again to spin-orbit interaction this causes a small decrease of the sample resistance.
 An applied magnetic field will destroy the edge spin polarization leading to a positive 
 magnetoresistance. This effect provides means to study spin accumulation by 
 electrical measurements. The origin and the general properties of the phenomenological 
 equations describing coupling between charge and spin currents are also discussed.

\end{abstract}

\maketitle

%----------------------------------------------------------------------------
It was predicted a long time ago \cite{dyakonov1,dyakonov2} that because of spin-orbit 
interaction electrical and spin currents are interconnected: an electrical current 
produces a transverse spin current and {\it vice versa}. In recent years this has
become a subject of considerable interest.

The purpose of this Letter is twofold. First, another way of understanding this 
interconnection will be presented and some general properties of the resulting 
phenomenological equations will be discussed. Second, a new magnetoresistance effect 
will be considered, which allows to study the current-induced spin accumulation near 
the sample edges by purely electric measurements.

The transport phenomena related to coupling of the spin and charge currents can be 
described phenomenologicaly in the following simple way. Let \text{\boldmath$q$} be
the electron flow density and let $\text{\boldmath$q$}^{(0)}$  be its conventional 
expression not accounting for spin-orbit interaction:  
$$\text{\boldmath$q^{(0)}$} = -\mu n \text{\boldmath$E$} - D\nabla n, \eqno{(1)}$$
where  $\mu$ and $D$ are the usual electron mobility and diffusion coefficient, 
connected by the Einstein relation, 
\text{\boldmath$E$} is the electric field, and  $n$ is the electron concentration. The
electric current density is $\text{\boldmath$j$}=-e\text{\boldmath$q$}$, where $e$ is the 
absolute value of the electron charge.  

Let $q_{ij}$ be the spin polarization current density tensor (the flow of the $j$ 
component of the spin polarization in the direction $i$). It should be understood that 
for polarized electrons the spin current may exist even in the absence of spin-orbit 
interaction, simply because spins are carried by electron flow. We denote the 
corresponding quantity as $q_{ij}^{(0)}$. Then, similar to Eq. (1), we have
$$q_{ij}^{(0)}=-\mu E_iP_j - D\frac {\partial P_j}{\partial x_i}, \eqno{(2)}$$
where \text{\boldmath$P$} is the vector of electron spin polarization density. If there 
are other sources for currents, like for example a temperature gradient, the corresponding 
terms should be included in Eqs. (1) and (2). 

We have departed from the conventional definitions
\cite{dyakonov1,dyakonov2} by introducing the vector of spin {\it polarization} density 
\text{\boldmath$P$} and the spin {\it polarization} current $q_{ij}$. This allows to avoid
numerous factors $1/2$ and $2$ in the formulas to follow. One can return to the
traditional notations by putting $\text{\boldmath$P$}=2\text{\boldmath$S$}$, where 
\text{\boldmath$S$} is the spin density, and by replacing $q_{ij}$ by $q_{ij}/2$ to
obtain the true spin current density.

Spin-orbit interaction couples the two currents. For a material with inversion symmetry 
\cite{rem1} we have:
$$q_i=q_{i}^{(0)}+\gamma \epsilon _{ijk}q_{jk}^{(0)}, \eqno{(3)}$$
$$q_{ij}=q_{jk}^{(0)}-\gamma \epsilon _{ijk}q_{k}^{(0)}, \eqno{(4)}$$
where $\epsilon _{ijk}$ is the unit antisymmetric tensor and $\gamma$ is a dimensionless 
coupling constant proportional to the spin-orbit interaction, it is assumed that 
$\gamma<<1$. The difference in signs in Eqs. (3) and (4) is consistent with the Onsager 
relations and is due to the different properties of \text{\boldmath$q$} and $q_{ij}$ 
with respect to time inversion \cite{rem2}.

Explicit phenomenological expressions for the two currents follow from Eqs. (1)-(4):
$$\text{\boldmath$j$}/e = \mu n \text{\boldmath$E$} +D\nabla n+
\beta \text{\boldmath$E$}\wedge \text{\boldmath$P$} +
\delta \,{\rm curl}\,\text{\boldmath$P$},\eqno{(5)}$$
$$q_{ij}=-\mu E_iP_j - D\frac {\partial P_j}{\partial x_i}+\epsilon _{ijk}(\beta nE_k + 
\delta \frac {\partial n}{\partial x_k}).\eqno{(6)}$$
Here 
$$\beta = \gamma \mu,\qquad \delta = \gamma D,\eqno{(7)} $$
so that the coefficients $\beta$ and $\delta$, similar to $\mu$ and $D$, satisfy
the Einstein relation.

Eqs. (5) and (6) should be complemented by the equation for the spin polarization 
vector:
$$\frac {\partial P_j}{\partial t}+\frac {\partial q_{ij}}{\partial x_i}+
(\text{\boldmath$\Omega$}\wedge \text{\boldmath$P$})_j+{\frac {P_j}{\tau_s}}=0,
\eqno{(8)}$$
where the vector \text{\boldmath$\Omega$} is directed along the applied magnetic field, 
$\Omega$ being the spin precession frequency and $\tau _s$ is the spin relaxation time. 
In Eqs. (6), (7) we ignore the action of magnetic field on the particle dynamics. This is 
justified if $\omega_c \tau<<1$, where $\omega_c$ is the cyclotron frequency and $\tau$ is 
the momentum relaxation time. Since normally $\tau_s >>\tau$, it is possible to have 
both $\Omega \tau_s>>1$ and $\omega_c \tau<<1$ in a certain interval of magnetic fields. 
It is also assumed that the equilibrium spin polarization in the applied magnetic field 
is negligible.

While Eqs. (5)-(8) are written for a three-dimensional sample, they are equally 
applicable to the 2D case, with obvious modifications: the electric field, space
gradients, and all currents (but not the spin polarization vector) should have 
components in the 2D plane only. 

In the equilibrium situation all currents should obviously vanish. If an inhomogeneous 
magnetic field exists, the equilibrium spin polarization 
will be space-dependent, however this by itself should produce neither spin, nor 
charge currents. To assure this, an additional counter-term should be introduced 
into the right-hand side of Eq. (2), proportional to ${\partial B_j}/{\partial x_i}$,
 which takes care of the force acting on the electron with a given spin in an 
 inhomogeneous magnetic field \text{\boldmath$B$}(\text{\boldmath$r$}) (see 
 \cite{liu}). Corresponding terms will appear in Eqs. (5), (6). We ignore these
 terms assuming that \text{\boldmath$B$} is homogeneous. 

Equations (5)-(8), which appeared for the first time in Refs. \cite{dyakonov1,dyakonov2} 
describe all the physical consequences of spin-charge current coupling \cite{rem3}. The term 
$\beta \text{\boldmath$E$}\wedge \text{\boldmath$P$}$ describes the anomalous Hall effect 
\cite{karplus}, where the spin polarization plays the role of the magnetic field. 

The term $\delta \,{\rm curl}\,\text{\boldmath$P$}$ describes an
electrical current induced by an inhomogeneous spin density (now referred to as the 
Inverse Spin Hall Effect). A way to measure this current under the conditions of optical 
spin orientation was proposed in \cite{averkiev}. The circularly
polarized exciting light is absorbed in a thin layer near the surface of the sample. As 
a consequence, the photo-created electron spin density is inhomogeneous, however 
$\,{\rm curl}\,\text{\boldmath$P$}=0$, since both \text{\boldmath$P$} and its gradient are
perpendicular to the surface. By applying a magnetic field parallel to the surface one can
create a parallel component of \text{\boldmath$P$}, thus inducing a non-zero 
$\,{\rm curl}\,\text{\boldmath$P$}$ and the corresponding surface electric current (or 
voltage). This effect was found experimentally for the first time by Bakun {\it et al}
\cite{bakun}. 

The term $\beta n\epsilon _{ijk}E_k$ (and its diffusive counterpart $\delta \epsilon _{ijk} 
{\partial n}/{\partial x_k}$) in Eq. (6), describes what is now called the Spin 
Hall Effect: an electrical current induces a transverse spin current, resulting in
spin accumulation near the sample boundaries \cite{dyakonov1,dyakonov2}. This phenomenon
was observed experimentally only in recent years \cite{kato,wunderlich} and has attracted 
widespread interest. 

It should be stressed that all these phenomena are closely related and have their common
origin in the coupling between spin and charge currents given by Eqs. (3) and (4). Any
mechanism that produces the anomalous Hall effect will also lead to the spin Hall effect
and {\it vice versa}. It is remarkable that there is a single dimensionless parameter, 
$\gamma$, that governs the resulting physics. The calculation of this 
parameter should be the objective of a microscopic theory. For the case, when the 
coupling is due to spin asymmetry in electron scattering, this was done in Ref. 
\cite{dyakonov2}, where $\beta$ and $\delta$ were expressed through the scattering amplitude. 
In this case $\gamma$ depends only on the form of the scattering potential, the 
electron energy, and the strength of spin-orbit interaction.

An "intrinsic" mechanism of the spin Hall effect, related only to spin band splitting, was 
proposed for bulk holes in the valence band \cite{murakami}. The value of $\gamma$ is on 
the order of $(k_F\ell)^{-1}$ ($k_F$ is the Fermi wavevector, $\ell$ is the mean free 
path) and generally depends on the details of the scattering mechanism. The current 
consensus is that the intrinsic mechanism may exist for any type of spin band splitting, 
{\it except} if it is linear in $k$ \cite{rem4}. 

Note that the $J=3/2$ holes may not be described by the simple 
Eqs. (5)-(8), because for higher spins the number of coupled macroscopic quantities
increases compared to spin 1/2 particles. The mutual transformation of spin and charge
currents for holes, due to scattering, was studied in \cite{khaetskii}. Also,  
even in the absence of spin-orbit interaction, holes are still particles with internal 
angular momentum $L=1$ and the splitting into light and heavy holes still exists. Thus, 
for the case of holes the spin-orbit interaction is not of primary importance.

We now discuss a new related phenomenon: a magnetoresistance due specifically to 
spin accumulation near the sample edges. Since the accumulation occurs on the scale 
of the spin diffusion length $L_s=\sqrt{D\tau_s}$ (the "spin layer" 
\cite{dyakonov1}), the proposed effect depends on the sample size, $L$, and 
becomes negligible when $L>>L_s$. 

Whithin the spin layer the $z$
component of spin polarization changes in the direction perpendicular to the sample
boundary (the $y$ direction). Thus $\,{\rm curl}\,\text{\boldmath$P$}\neq 0$, and
according to Eq. (5) a correction to the electric current should exist. As we will 
see, this correction is positive, i.e. it leads to a slight decrease of the sample 
resistance compared to the (hypothetical) case when spin-orbit interaction is absent.
 By applying a magnetic field in the $xy$ plane, we can destroy the spin polarization 
 (the Hanle effect) and thus observe a positive magnetoresistance on a field scale 
corresponding to $\Omega \tau_s \sim 1$. One might say that this is a manifestation 
of combined direct and inverse spin Hall effects, and the Hanle effect.

  %______________________ Fig.1___________________________________
\begin{figure}
\epsfxsize=235pt {\epsffile{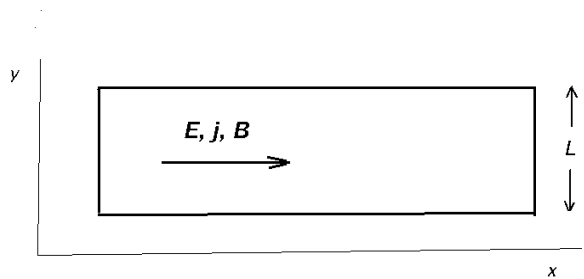}}
\caption{The geometry of the proposed experiment with a 2D sample. The direction of the 
magnetic field is of no importance, so long as it lies in the 2D plane.
For simplicity it is assumed to be parallel to the current.}
\end{figure}
%________________________________________________________________%

We will consider a 2D sample (see Fig. 1), a similar effect will exist also for a thin
wire. The advantage of the 2D case is that the small effect considered here will not be 
masked by the normal magnetoresistance, because the magnetic field parallel to the 2D
plane acts on the spins only, but not on the electron orbital motion. Since the spin 
polarization is proportional to the electric field, we discard nonlinear in $E$ 
terms proportional to $EP$. 

For the geometry of Fig. 1, from Eq. (5) we obtain:$$j=e(\mu n E + \delta {\frac {dP_z}{dy}}),
\eqno{(9)} $$ so that the total current is $$I=\int_{-L/2}^{L/2}j(y)dy=I_0+\Delta I,\eqno{(10)}$$
where $$I_0=e\mu n EL,\quad \Delta I=e\delta(P_z(L/2)-P_z(-L/2)). \eqno{(11)}$$
The correction to the current, $\Delta I$, is proportional to the difference in 
spin polarization at the opposite edges of the sample. Eq. (6) yealds:
$$q_{yz}=-D{\frac {dP_z}{dy}}+\beta nE,\qquad q_{yy}=-D{\frac {dP_y}{dy}}.\eqno{(12)}$$
In the steady state Eq. (8) gives:
$$D{\frac {d^2P_z}{dy^2}}-\Omega P_y ={\frac {P_z}{\tau_s}},\quad D{\frac
{d^2P_y}{dy^2}}+\Omega P_z ={\frac {P_y}{\tau_s}}.\eqno{(13)}$$
These equations should be solved with the boundary conditions at $y=\pm L/2$:
$${\frac {dP_z}{dy}}={\frac {\beta nE}{D}}, \qquad {\frac {dP_y}{dy}}=0, \eqno{(14)}$$
corresponding to vanishing spin currents $q_{yz}$ and $q_{yy}$ at the sample edges.

%______________________ Fig.2___________________________________
\begin{figure}
\epsfxsize=235pt {\epsffile{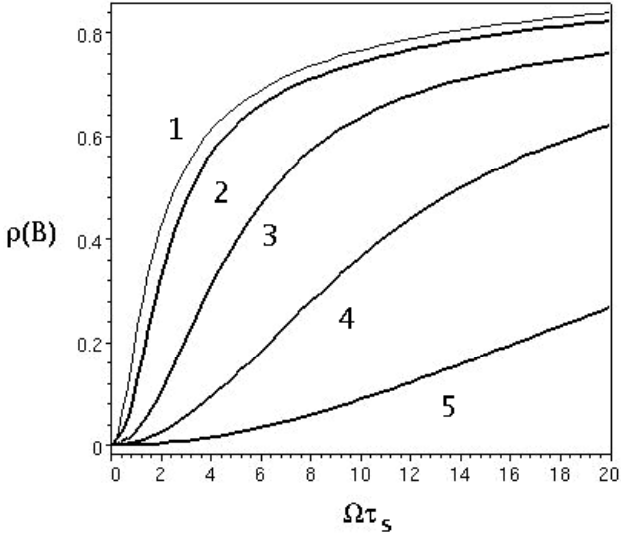}}
\caption{Normalized magnetoresistance as a function of the parameter $\Omega \tau_s$ 
for different sample widths. {\it 1} - Eq. (18) for $\lambda >>1$, 
 {\it 2} - $\lambda =1.5$, {\it 3} - $\lambda =0.8$, {\it 4} - $\lambda =0.5$, 
 {\it 5} - $\lambda =0.3$}.
\end{figure}
%________________________________________________________________%

A straightforward calculation gives the result:
$${\frac {\Delta R}{R_0}}=-{\frac {\Delta I}{I_0}}= -\gamma ^2\text {Re}\Bigl 
[{\frac {\tanh (\kappa \lambda)}{\kappa \lambda}}\Bigr ], \eqno{(15)}$$
where $R_0$ is the uncorrected sample resistance, $\Delta R$ is the field-dependent 
negative correction due to spin accumulation,  
$$\kappa = (1- ix)^{1/2}, \qquad x=\Omega \tau_s, \qquad \lambda = L/(2L_s),$$ 
and $L_s=(D\tau_s)^{1/2}$ is the spin diffusion length. 

Thus $\Delta R$ is proportional to the square of the dimensionless parameter $\gamma$ in 
Eqs. (3), (4). In deriving Eq. (15) the relations given by Eq. (7) were used. 

From this result one can easily deduce two characteristic features of this effect.

1) {\it The total resistance change} between its zero-field value, $R(0)$ and its value 
at strong enough field ($\Omega \tau_s>>1)$, $R(\infty)=R_0$:
$${\frac {R(\infty)-R(0)}{R(0)}}=\gamma ^2 {\frac {\tanh\lambda}{\lambda}}, 
\eqno{(16)}.$$
For a narrow sample, $\lambda <<1$ the overall relative change of resistance is equal to 
$\gamma ^2$, which gives a nice way to determine experimentally the fundamental 
parameter $\gamma$. For wide samples the relative change is $\gamma ^2 (2L_s/L)$.

2){\it The shape of the magnetoresistance curve}. We introduce the notation $\rho (B)$ 
for the normalized relative magnetoresistance. Then
$$\rho(B)={\frac {R(B)-R(0)}{R(\infty)-R(0)}}=1- \text {Re}\Bigl 
[{\frac {\tanh (\kappa \lambda)}{\kappa \tanh \lambda}}\Bigr ], \eqno{(17)}$$

For a wide sample, $\lambda>>1$, the width of the magnetoresistance curve is determined 
by the condition $\Omega \tau_s \sim 1)$. In this case 
$$ \rho(B)=1-\text {Re} \Bigl ({\frac {1}{\kappa}} \Bigr )=1-\Bigl [{\frac {1+\sqrt{1+x^2}}
{2(1+x^2)}} \Bigr ]^{1/2}.\eqno{(18)} $$
Note that at $x=\Omega \tau_s >>1$ the function $\rho(B)$ approaches its maximum value  
very slowly, as $1-1/\sqrt {2x}$.

Figure 2 presents numerical results for $\rho(B)$ calculated from Eq. (17) for different 
values of $\lambda = L/(2L_s)$ together with the curve given by Eq. (18) for $\lambda>>1$, 
which is a good approximation already for $\lambda=1.5$. 

For narrow samples ($\lambda<<1$) the magnetic field dependence becomes much weaker. The 
reason is that along with the spin relaxation time $\tau_s$, there is another characteristic
time, $\tau_d=\tau_s\lambda^2 =L^2/(4D)$, which is the time of diffusion on a distance $L/2$. 
For narrow samples ($\tau_d<\tau_s$), it is this time, rather than $\tau_s$, that 
determines the width of the Hanle curve, because the spin polarization is destroyed by 
diffusion faster than by spin relaxation. Accordingly, the width of the magnetoresistance 
curve will now correspond to $\Omega \tau_d \sim 1$, i.e. it will be $1/\lambda^2$ times 
broader compared to the case of a wide sample.

%______________________ Fig.3___________________________________
\begin{figure}
\epsfxsize=235pt {\epsffile{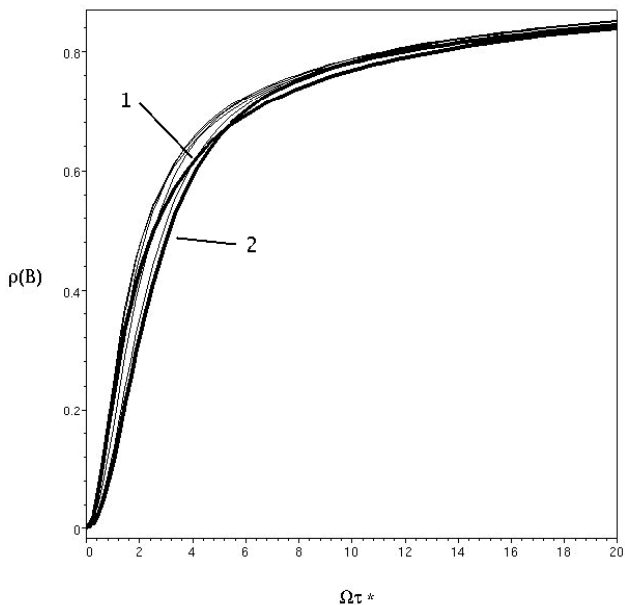}}
\caption{Normalized magnetoresistance as a function of the parameter $\Omega \tau^*$,
where $1/\tau^*= 1/\tau_s+1/\tau_d$. {\it 1} - in the limit $\lambda>>1$, {\it 2} - in 
the limit $\lambda<<1$. Other curves correspond to $\lambda = 0.4$, 0.8, 1.2, 1.6, and 2.
One can see that all curves practically coincide.}
\end{figure}
%________________________________________________________________%

To unify the two limiting cases, in Fig. 3 we re-plot $\rho(B)$ as a function of the 
parameter 
$\Omega \tau^*$, where $\tau^*$ is the effective time during which the spin is destroyed 
because of the combined effect of spin diffusion and spin relaxation:
$${\frac {1}{\tau^*}}= {\frac {1}{\tau_s}}+{\frac {1}{\tau_d}}={\frac {1}{\tau_s}}(1+
{\frac {1}{\lambda^2}}).$$
Figure 3 shows that as a function of this parameter there is a quasi-universal curve, since 
the results for the limiting cases of a narrow and a wide samples are very close. Thus Eq.
(18) can serve as a good interpolation formula for the general case, provided the variable 
$x$ is replaced by $\Omega \tau^*$, instead of $\Omega \tau_s$. The high-field limit is 
always approached as $1/\sqrt{B}$.

The above results for magnetoresistance are similar to those obtained previously 
\cite{dyakonov3} for the Hanle effect in the case when spin polarization is inhomogeneous 
and spin diffusion is important. 

From the experimental results for 3D \cite{kato} and 2D \cite{liu} GaAs one can estimate 
$\gamma \sim 10^{-2}$. Kimura {\it et al} \cite{kimura} find $\gamma =3.7\cdot 10^{-3}$ for 
Platinum at room 
temperature, so that in these two cases a magnetoresistance due to spin accumulation on the 
order of $10^{-4}$ 
and $10^{-5}$, respectively, can be expected. The characteristic feature, allowing to 
identify this effect, is the specific form of the magnetoresistance curve, as well as its 
strong dependence  on the sample width, when it becomes comparable to the spin diffusion 
length. 

Because of the high precision of electrical measurements, magnetoresistance might 
provide a useful tool for studying the spin-charge interplay in semiconductors and metals.


\begin{thebibliography}{}
\bibitem{rem1} In the absence of inversion symmetry there may be additional terms 
describing this coupling. In particular, there is a spin current induced by a non-equilibrium 
spin polarization and a uniform spin polarization generated by electric current.
\bibitem{rem2} A simple way to verify these equations is to start with  
currents $\text{\boldmath$q$}^{\pm}$ for particles with spin-up and spin-down (with 
respect to the $z$ axis), assuming that the electric field and the concentration 
gradient are along $x$: $q_{x}^{\pm} = -\mu n_{\pm} E - D{\partial n_{\pm}}/{\partial x}$.
(Spin-up and spin-down mobilities are taken equal, which is true when the 
spin polarization is small). Because of spin-orbit interaction these currents will induce 
currents of opposite 
signs in the $y$-direction: $q_{y}^{\pm}= \mp \gamma q_{x}^{\pm}$. It can now be seen that 
the expressions for $q_y = q_{y}^{+}+q_{y}^{-}$ and $q_{yz}=q_{y}^{+}-q_{y}^{-}$ 
coincide with what is given by Eqs. (1)-(4) with $P_z(x)=n_+ - n_-$. If, instead of the 
spin {\it polarization} current, the normal spin current were used, we would have to 
replace the coefficient $\gamma$ by $2\gamma$ in Eq.(3) and by $\gamma /2$ in Eq.(4).
\bibitem{rem3} In \cite{dyakonov1,dyakonov2} these equations were written in terms of 
spin density \text{\boldmath$S$}. Thus the definitions of the 
coefficients in Eqs. (5), (6), and the relations between them, were different (by 
numerical factors) from those adopted here.
\bibitem{rem4} However, it does not seem reasonable to introduce "extrinsic" and "intrinsic" 
 spin Hall effects as distinct physical phenomena. The phenomenon is always the same: current
 induced edge spin accumulation, while certainly there may be different microscopic mechanisms 
 for spin-charge current coupling.
\end{thebibliography}
\end{document}